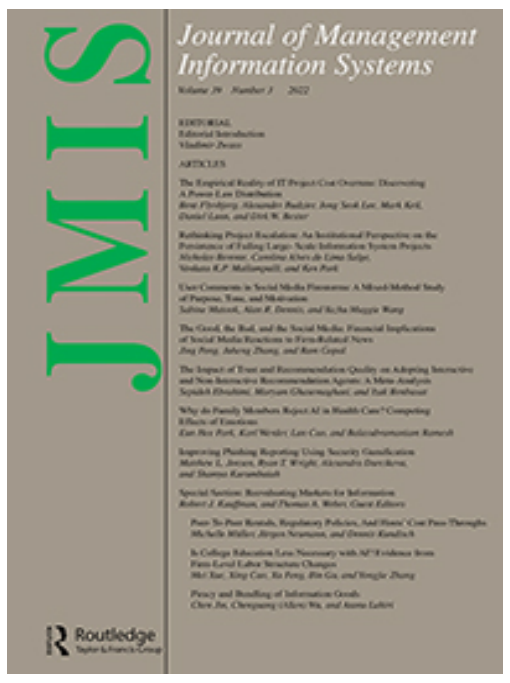

# The Empirical Reality of IT Project Cost Overruns: Discovering A Power-Law Distribution

Bent Flyvbjerg, Alexander Budzier, Jong Seok Lee, Mark Keil, Daniel Lunn & Dirk W. Bester

# The Empirical Reality of IT Project Cost Overruns: Discovering A Power-Law Distribution

Bent Flyvbjerg[a,b], Alexander Budzier[c], Jong Seok Lee[d], Mark Keil[e], Daniel Lunn[f], and Dirk W. Bester[g]

[a]First BT Professor and Inaugural Chair of Major Programme Management, University of Oxford's Saïd Business School; [b]Villum Kann Rasmussen Professor and Chair of Major Program Management, IT University of Copenhagen; [c]Fellow in Management Practice in the Field of Information Systems, Saïd Business School, University of Oxford; [d]Department of Accounting and Information Management, Haslam College of Business, University of Tennessee, Knoxville; [e]Regents' Professor of the University System of Georgia, John B. Zellars Professor of Computer Information Systems, J. Mack Robinson College of Business, Georgia State University; [f]Department of Statistics, University of Oxford; [g]Independent Researcher, 4 Thornhill Mews Cross Street, Maidstone, Kent, ME14 2SP, UK

**ABSTRACT**

If managers assume a normal or near-normal distribution of Information Technology (IT) project cost overruns, as is common, and cost overruns can be shown to follow a power-law distribution, managers may be unwittingly exposing their organizations to extreme risk by severely underestimating the probability of large cost overruns. In this research, we collect and analyze a large sample comprised of 5,392 IT projects to empirically examine the probability distribution of IT project cost overruns. Further, we propose and examine a mechanism that can explain such a distribution. Our results reveal that IT projects are far riskier in terms of cost than normally assumed by decision makers and scholars. Specifically, we found that IT project cost overruns follow a power-law distribution in which there are a large number of projects with relatively small overruns and a fat tail that includes a smaller number of projects with extreme overruns. A possible generative mechanism for the identified power-law distribution is found in interdependencies among technological components in IT systems. We propose and demonstrate, through computer simulation, that a problem in a single technological component can lead to chain reactions in which other interdependent components are affected, causing substantial overruns. What the power law tells us is that extreme IT project cost overruns will occur and that the prevalence of these will be grossly underestimated if managers assume that overruns follow a normal or near-normal distribution. This underscores the importance of realistically assessing and mitigating the cost risk of new IT projects up front.

## Introduction

Undertaking large Information Technology (IT) projects can be risky. Both Kmart, a large U.S. retailer, and Auto Windscreens, a major U.K. automobile glass company, were driven into bankruptcy in part due to their inability to manage large IT projects [30]. Project failure

**CONTACT** Bent Flyvbjerg flyvbjerg@mac.com First BT Professor and Inaugural Chair of Major Programme Management, University of Oxford's Saïd Business School; Villum Kann Rasmussen Professor and Chair of Major Program Management, IT University of Copenhagen has also been known to ruin the careers of executives. U.S. Health Secretary Kathleen Sebelius [64] and TSB Bank CEO Paul Pester [57] both lost their jobs due to mismanaged IT projects.

The extent to which a project is completed within budget constitutes a key dimension of project performance [43, 59, 71]. Moreover, cost overruns can have an obvious negative impact on an organization's ability to achieve a positive return on investment (ROI). Nonetheless, industry reports and academic studies alike suggest that it is quite common for IT projects to experience cost overruns [14, 39]. The U.S. Department of Defense reports that for the fiscal year ending in 2020, IT project spending was $37 billion and only 35% of the projects were within budget [17]. A study by McKinsey and the BT Centre for Major Programme Management at the University of Oxford reports that on average, large IT projects run 45 percent over budget [14]. What is worth noting in these reports and studies is that some IT projects have very large cost overruns of around 200% [30, 39] or even 400% [14]. From a statistical point of view, based on these numbers it is reasonable to ask whether the probability distribution of IT project overruns follows a Gaussian or near-Gaussian distribution (i.e., normal or near-normal distribution), as often assumed, or if it instead follows a power-law distribution. If managers assume a normal distribution of IT project cost overruns, as is common, and cost overruns can be shown to follow a power-law distribution, managers may be unwittingly exposing their organizations to extreme risk by severely underestimating the probability of large cost overruns. Unfortunately, little is known about the empirical distribution of IT project cost overruns. Therefore, in this research, we aim to address two research questions:

(RQ1) What is the probability distribution of cost overruns in IT projects?

(RQ2) How can we explain the probability distribution of cost overruns in IT projects?

A power-law distribution is characterized by having a big head (i.e., many small values clustered at the top of the distribution) and a fat right tail (i.e., a greater number of large values relative to what one might expect from a normal distribution) [19]. For example, in general the distribution of income is said to follow a power-law distribution; the vast majority of the population make small and modest incomes, but there are many billionaires in the tail of the distribution [20]. In the context of IT project cost overruns, a power-law distribution would entail a large number of projects with relatively small overruns and a fat tail that includes a smaller number of projects with very large overruns. Unfortunately, extant research offers limited insight into the probability distribution of IT project cost overruns. This is an important issue, because misunderstanding the probability distribution of cost overruns may lead decision makers to underestimate or ignore the very real risk of experiencing substantial cost overruns on their IT projects. Research suggests that when low-probability but high-impact events are ignored, decision makers tend to make risky [33] and suboptimal decisions [24]. When it comes to IT project decisions, this can mean undertaking highly risky projects [38], ignoring project risks [47], or neglecting testing and quality assurance [4]. Therefore, shedding light on the actual probability distribution of IT project cost overruns can help decision makers avoid making decisions that could ultimately prove to be quite costly.

This research assembles a large sample comprised of 5,392 IT projects completed between 2002 and 2014. The total cost of the IT projects in the sample was $56.5 billion measured in 2015 USD. From the outset, our goal was to assemble a large sample, because samples in prior studies were too small to render a clear understanding of the probability distribution of IT project cost overruns. Small samples suffer from statistical noise, can be unreliable, and are not representative of the population. Our approach in this research was to first explore the probability distribution of IT project cost overruns using a large sample of IT projects to address RQ1 and this led to discovering a power-law distribution. Next, drawing on both IT project management research and literature on the generative mechanisms for power-law distributions, we propose a theoretical explanation for the power-law distribution observed in our data to address RQ2. Specifically, we draw on the concepts of self-organized criticality and interdependencies among technological components in an IT system to propose a generative mechanism for the power-law distribution of IT project cost overruns. Finally, we conducted a computer simulation to test the proposed generative mechanism.

This research makes several important contributions that advance our understanding of IT project management. First, we uncover a power-law distribution that describes IT project performance, which has never been done before and represents an empirical, phenomenological law of a generic nature, as described by Mandelbrot [50]. Our findings underscore the fact that IT projects may be far riskier than normally assumed. When the normal distribution is assumed, this can cause managers to substantially underestimate the risk associated with IT projects. Second, we propose a generative mechanism that may explain the power-law distribution in IT project cost overruns.

## Background

### IT Project Performance and Overruns

We conducted a literature review to assess extant knowledge on IT project performance (see Appendix A for the detailed procedure of the review). Our review revealed that most of the prior research focused on factors influencing IT project performance, and identified such causal factors as project coordination [59], project risks [31], and project governance-knowledge fit [71]. In addition, our review revealed that researchers have assessed IT project performance using both objective measures (e.g., actual expenditures) [63] and subjective measures (e.g., by asking managers to provide the percentage of an overrun in comparison to the project's budget) [71].[1] Most importantly, our review revealed that there has been no attempt to examine the probability distribution of IT project cost overruns. Table 1 provides a list of studies that examined and measured IT project overrun as a proportion (or percentage) of original budget, schedule, or effort.

[1]In the literature, IT project success or performance has been typically assessed in two ways: the extent to which a project is completed within budget and schedule (process performance) and the extent to which the delivered system meets the needs of its intended users (product performance) [59]. In this research, we focus on process performance as our goal is to understand the probability distribution of cost overruns. In addition, while process performance such as cost overruns can be measured objectively using archival data, product performance involves subjective evaluation of the system by one or more stakeholder groups and is typically measured using a survey instrument. Surveys tend to employ fixed scales and are not well suited for capturing extremes values. Therefore, to empirically examine if project performance follows a power-law distribution, it is appropriate to focus on process performance (e.g., project cost performance).

**Table 1.** Representative Studies that Measured IT Project Performance

| Study | Data source | Measure | Sample | Mean | SD | Max |
|---|---|---|---|---|---|---|
| Governance-knowledge fit in systems development projects [71] | Survey responses from MIS directors | Percentage overrun relative to the originally planned level in the project's budget, schedule, and development/ programming effort (person-months) | 89 internal application development projects from 89 firms | Cost overrun: 10.2% Schedule overrun: 23.4% Effort overrun: 15.9% | Cost overrun: 46.18% Schedule overrun: 58.9% Effort overrun: 50.7% | NR* |
| Offshore information systems project success: the role of social embeddedness and cultural characteristics [63] | Archival Project Data | The percentage difference between actual project costs and budgeted project costs. | 155 offshore IS projects from a software vendor in India | Cost overrun: 25.28% | Cost overrun: 10.87 | NR |
| An integrative contingency model of software project risk management [10] | Survey responses from project managers | Cost Gap = 1 – (actual cost of project/ estimated cost of project) | 75 software development projects from multiple organizations | -0.17 | 0.75 | -2.47 |
| How user risk and requirements risk moderate the effects of formal and informal control on the process performance of IT projects [40] | Survey responses from user liaisons who were responsible for overseeing the project from a user perspective | The extent to which the project was completed within budget and schedule (5-point Likert scale) | 63 IT projects from multiple companies | 3.43 | 0.70 | NR |
| The effects of user partnering and user non-support on project performance [36] | Survey responses from IS managers or project managers | Adherence to budget and schedule (5-point Likert scale) | 170 Project Management Institute (PMI) members were asked to provide information about the most recently completed IS project | 3.56 | .75 | NR |
| Estimation of information systems development efforts [13] | Survey responses from software development managers | Estimate accuracy = (actual effort - estimated effort) /estimated effort Effort is defined as person-days | 89 software development projects from 63 organizations | 33% | 31% | 100% |

(*Continued*)

**Table 1.** (Continued).

| Study | Data source | Measure | Sample | Mean | SD | Max |
|---|---|---|---|---|---|---|
| Schedule estimation and uncertainty surrounding the cone of uncertainty [48] | Archival project data | Actual project duration / estimated project duration | 106 software development projects in a company | 2.0 | NR | $6.7^{+}$ |
| Successful development strategies for business application systems [53] | Survey responses from project leads | Percentage overrun relative to the originally planned level in the project's budget, schedule, and effort (manhours) | 32 application systems from 5 organizations | Cost overrun: 28%<br>Schedule overrun: 22%<br>Effort overrun: 25% | NR | NR |
| The effects of coupling IT and work process strategies in redesign projects [54] | Interviews with project managers | Schedule Slippage (additional months taken to complete a project) | 43 IT projects in an organization | 9.73 (months) | 9.03 (months) | NR |
| Empirical investigation of systems development practices and results [34] | Interviews with systems development managers | Estimated – actual/estimated | 72 systems development projects from 23 organizations | Cost overrun: 66.7%<br>Schedule overrun: 22%<br>Effort overrun: 36% (person-days) | NR | Cost overrun: 525.7%<br>Schedule overrun: 525%<br>Effort overrun: 525% (person-days) |
| Detection of early warning signals for overruns in IS projects: Linguistic analysis of business case language [12] | Archival project data | Percentage overrun relative to the originally planned level in the project's budget and schedule | 6 Dutch governmental information systems projects | One experienced no budget overrun; one experienced an overrun of less than 100%; two experienced an overrun of between 100% and 200%; two experienced an overrun of more than 200%.<br>One experienced no schedule overrun; four experienced an overrun of less than 100%; one experienced an overrun of between 100% and 200% | NR | NR |
| Project failure en masse: A study of loose budgetary control in ISD projects [22] | Interviews with managers | A deviation from the original budget | 4 information systems projects in a business unit | Budget deviation for each project: 12%, 4%, 223%, 320% | NR | NR |

(*Continued*)

**Table 1.** (Continued).

| Study | Data source | Measure | Sample | Mean | SD | Max |
|---|---|---|---|---|---|---|
| The role of organizational controls and boundary spanning in software development outsourcing: Implications for project performance [32] | Survey responses | Project's overruns on schedule, person-month effort, and cost as a proportion (overrun measures are multiplied by −1 (reverse coded) to indicate that numerically higher values represent smaller overruns | 96 software projects from 10 firms | Overall overruns of schedule, effort, cost<br>Mean: −17.118 | Overall overruns of schedule, effort, cost<br>SD: −20.578 | NR |
| A comparison of software project overruns-flexible versus sequential development models [56] | Interviews with managers | Deviations between estimated effort/schedule and actual effort/schedule | 42 software projects | Effort:<br>Flexible development method (mean 0.09; median 0.01)<br>Sequential development method (mean 0.29; median 0.38)<br>Schedule:<br>Flexible development method (mean 0.10; median 0.06)<br>Sequential development method (mean 0.15; median 0.10) | NR | NR |

*NR = not reported
+This is an approximate number derived from the probability distribution provided in the paper.

While most of these studies tend to report only means and standard deviations of project overruns, there are a few studies that reported maximum values (e.g., the largest cost overruns). Maximum values are more informative for understanding the probability distribution of IT project overruns and examining the potential presence of a fat tail. Little] investigated the performance of 106 software development projects and found that on average it took twice as long to complete a project than originally estimated. He found that the most extreme case was close to a 700% overrun. Moreover, he examined the cumulative distribution of the schedule overrun data (the ratio of actual to estimated schedule) and concluded that the data followed a lognormal distribution rather than a normal distribution. These findings offer valuable insight suggesting that IT project performance may not follow a normal distribution. However, Little's [48] study was limited in the sense that it was based on a small sample (e.g., 106 projects) from a single company. Similar to Little [48], there are other studies that have reported fairly large overruns. For example, based on interviews with systems development managers Jenkins et al. [34] obtained IT project performance data on 72 systems development projects from 23 organizations and found that the largest cost overrun was 525%. In addition, Langer et al. [45] examined the actual costs of 530 projects conducted by a software vendor in India and found that this data was skewed (i.e., not normally distributed). Therefore, they used the log-transformed variable in their analysis.

If IT project performance does not follow a normal distribution, it could have important implications for both IT project management research and practice, and further research is warranted to understand the actual form of the distribution. To advance our understanding in this area, a large sample study of IT project performance is needed.[2] In this research, we collected such a dataset and examined the cost performance of IT projects (i.e., cost overruns). Cost performance is recognized as an important element of IT project performance because when the cost of a project becomes too high or excessive relative to a planned budget it reduces the net value of the project [59]. Further, since cost performance tends to be highly correlated with schedule performance, projects that exceed their budgets typically also exceed their schedule. While our primary focus was on understanding the probability distribution of IT project cost overruns, we also examined the probability distribution of IT project schedule and effort overruns for a subset of the projects in our sample, thus providing empirical evidence that not only cost but also schedule and effort overruns follow a power-law distribution.

### Interdependency in IT Systems

An IT project involves developing and deploying an IT system in an organization. Further, a modern technical system such as an IT system [21] can be understood as a system of systems in which a collection of systems or technologies (i.e., technological components) must be made to interoperate together in order for the system as whole to achieve desired objectives [1, 23]. For example, the Internet of Things (IoT) is "a system of wireless, interrelated, and connected digital devices that can collect, send, and store data over a network without requiring human-to-human or human-to-computer interaction." [41, p. e20135] In recent years, the IoT has created new opportunities for businesses and organizations. In the healthcare industry, government leaders and decision makers are encouraging the use of emerging IoT technologies to enable flexible and remote modes of care in response to the COVID-19 pandemic [41]. Specifically, in South America where severe difficulties have been encountered due to the shortage of medical resources and the need for social distancing, several countries have implemented home monitoring smart devices and sharable Electronic Health Records (EHR), enabling seamless exchange of medical data not only between patients and healthcare providers but among healthcare providers as well as administrative entities [18]. These smart devices and EHRs can be regarded as intelligent agents that create a network of systems processing, collecting and analyzing data, and transferring data to different entities [26].

[2]While there is one IS study that involved a sample of 2,378 open-source projects hosted at SourceForge [67], open-source projects are different from IT projects undertaken in organizations. Moreover, this study did not examine traditional measures of project performance (e.g., cost performance), but instead focused on the rate of knowledge creation in an open-source project.

The basic notion that an IT system is comprised of multiple technological components, such as self-contained subsystems and software applications, that must work together to form an overarching system is well-accepted in the IS literature (see, for example, Tiwana [70]). Moreover, IS researchers have highlighted the difficulty of building IT systems due to the interdependencies among technological components. Specifically, research suggests that there are inherent interdependencies among modules in a system and that such interdependencies can cause delays and errors during implementation, testing and integration phases of a project [27]. Further, research shows that decoupling system modules (i.e., reducing interdependencies among system modules) can help mitigate risk in digitization projects [75].

In sum, building IT systems in which underlying technological components are interdependent requires substantial effort to make these components work together as a whole and a failure in a single component can have impact on other interconnected components. When components are unconnected, a problem in a single component is isolated from the rest of the system. However, in a system consisting of interconnected components that may include software, sensors, and communication devices (like in the IoT), a problem in a single component can lead to chain reactions affecting other connected components [46]. In this research, we focus on interdependencies among technological components [8, 21, 72] in an IT system in explaining substantial overruns in IT projects which can shape the probability distribution of IT project cost overruns.

## An Empirical Investigation of IT Project Cost Overruns

### *Data Sources and Procedures*

To empirically examine the probability distribution of IT project cost overruns, we collected data on IT project performance with a focus on estimated project costs and actual project costs. We also collected data on estimated and actual schedule/effort when such data was available. We relied on two sources for our data collection: (1) primary data collection from firms willing to share their project data and (2) primary data collection through U.S. government freedom of information requests (i.e., U.S. government project data).

First, we used our professional contacts to identify firms willing to share their IT project data for this research. In addition, we reached out directly (i.e., without leads from our contacts) to firms and solicited their participation. The companies we contacted for data ranged from financial services, which typically undertake a large number of IT projects, to manufacturing, which have comparatively few projects. In total, we contacted 367 firms, of which 29 agreed to participate and provide data.[3] Some of the 29 organizations were software solution providers or IS consulting firms and provided data on projects that were undertaken by multiple client organizations. For each firm identified through these procedures, we probed to determine if the firm documented key project data, such as estimated costs and whether it recorded actual project performance including actual costs. We only included data from firms that were keeping track of estimated and actual project performance in a formal manner.[4] Each firm provided data about projects that were completed within the last five years at the time of the data collection.[5] We included all types and sizes of IT projects in order to avoid potential biases that can be associated with restricting the sample to particular types of projects or artificial cut-off values [37]. Our data collection was also independently audited by a postdoctoral researcher who examined our data collection procedures for accuracy and reliability. Following this approach, we were able to collect data on 2,739 IT projects that were undertaken by 872 different firms.

Second, U.S. federal government agencies undertaking a capital investment project under the 1996 Clinger-Cohen Act must submit a budget request form (named Exhibit 300, or E300) to the U.S. Office of Management and Budget (OMB). In addition, agencies are required to submit a budget request form (named Exhibit 53, or E53) for IT projects on an annual basis. With this form, agencies must provide project descriptions as well as detailed cost estimates. Further, they must report actual costs and updated forecasts for the approved projects for each fiscal year. We were able to access these project data by U.S. government agencies through U.S. government freedom of information requests. The Freedom of Information Act (FOIA) is a federal freedom of information law that requires the disclosure of information and documents controlled by the United States government upon request. Through the FOIA requests, we obtained data on 2,653 IT projects undertaken by 104 U.S. government agencies.

Taken together, we collected data on 5,392 IT projects that were completed between 2002 and 2014.[6] Among those, we were able to obtain both estimated and actual cost data on 4,677 projects. Our sample of projects is quite diverse, spanning over 66 countries in 6 continents and covering different types of IT projects undertaken by both the private and public sectors.[7] The total amount spent on the projects in the sample is USD 56.5 billion in 2015 prices.

[3]We note that firms that care to understand IT project performance, track project performance (e.g., cost overruns), and are transparent about project performance may be more likely to share data on IT projects. The implication here is that IT projects in these firms exhibit better performance in general than those in firms that are unwilling to share data. From this perspective, project performance in companies not included in our research may be worse than project performance in companies included in our research. Thus, cost overruns could be worse in the entire population than what is observed in our research.

[4]We note that projects that were terminated were not included in our data, as organizations tend not to track or are unable to track final performance data of such projects. Nonetheless, projects are terminated when they experience significant problems and including such projects would make performances data worse not better.

[5]The data we attempted to collect included project name, type, scope description, country, client, estimated cost, schedule, effort, and benefits at final investment decision, as well as actual cost, schedule, effort, and benefits. For the cost we also collected information on currency and year of the price level. Further we added information about industry and sector based on available knowledge about the firm.

[6]In the 13 years that the data cover the size of cost overruns did not change in any statistically significant manner ($p = 0.737$, linear regression on log-transformed overrun).

[7]Our data includes a large number of government projects whose performances may differ from those in the non-government context. Nonetheless, we found no statistically significant difference when we compared the power-law fit between data from the U.S. government and the data obtained from participating firms.

**Table 2.** Descriptive Cost Statistics of IT Projects

| | Estimated cost | Actual cost | Cost overrun |
|---|---|---|---|
| Mean | 6.6 | 14.7 | 1.8 |
| Median | 0.7 | 0.9 | 1.0 |
| Sd | 17.6 | 170.2 | 8.5 |
| Min | 0.0005 | 0.0002 | 0.0014 |
| Max | 593.5 | 8,676.7 | 280.4 |
| N | 4,627 | 3,835 | 4,677 |

Notes:
1. Very small costs of less than USD 1,000 were for projects in India and Mexico. These projects were all carried out at a very low rate per person day for development
2. The largest cost overrun was on a small workflow customization project, that initially was budgeted at USD 1,500 and ended up costing USD 425,000 (in 2015 prices)
3. The sample sizes (N) for estimated and actual cost are smaller that 4,677 due to inability to convert some costs to 2015 dollars for multinational projects. We did our power-law fitting, however, based on the 4,677 figure for cost overruns.
4. Estimated and actual costs are shown in USD in millions in 2015 prices. Cost overrun is actual cost divided by estimated cost.

## *Descriptive Statistics*

Table 2 provides descriptive cost statistics for our sample of IT projects. The mean of the estimated costs in our sample was USD 6.6 million (in 2015 prices).[8] The mean of the actual costs in our sample was USD 14.7 million (in 2015 prices). We also computed the cost overrun for each project by dividing the actual cost[9] by the estimated cost (we used the values in the original currencies for this calculation). The mean of the cost overruns expressed as a ratio of actual/estimated cost was 1.8. Table 3 shows descriptive cost statistics by project type and Table 4 shows descriptive cost statistics by public vs. private sector.

Next, we visually examined the probability distribution of cost overruns for the projects in our sample (Figure 1). The probability distribution appeared to be non-Gaussian, possibly with a fat tail, with a mode around 0% (ratio of 1.0). These observations run counter to those of earlier studies which suggested that cost overruns tend to be more common than underruns [13, 34, 48, 54, 60]. Contrary to this, our data indicate that overruns and underruns are about equally frequent.

**Table 3.** Descriptive Statistics by Project Type[10]

| | Estimated cost | | | | | Actual cost | | | | | Cost overrun | | | | |
|---|---|---|---|---|---|---|---|---|---|---|---|---|---|---|---|
| | ERP | HRM | MIS | SCM | Other | ERP | HRM | MIS | SCM | Other | ERP | HRM | MIS | SCM | Other |
| Mean | 3.9 | 6.0 | 5.7 | 7.5 | 9.4 | 12.5 | 8.4 | 9.3 | 4.4 | 26.4 | 1.3 | 1.5 | 1.8 | 2.2 | 2.2 |
| Median | 0.5 | 1.0 | 0.9 | 0.3 | 1.7 | 2.1 | 1.6 | 1.2 | 0.2 | 2.9 | 0.9 | 1.0 | 1.0 | 1.0 | 1.0 |
| Sd | 17.5 | 11.3 | 9.6 | 20.6 | 18.1 | 76.8 | 34.4 | 24.4 | 19.8 | 270.1 | 3.0 | 2.6 | 5.5 | 13.0 | 10.8 |
| Min | 0.001 | 0.001 | 0.001 | 0.001 | 0.001 | 0.002 | 0.001 | 0.001 | 0.001 | 0.002 | 0.003 | 0.002 | 0.078 | 0.014 | 0.001 |
| Max | 593.5 | 44.3 | 44.3 | 215.9 | 203.4 | 1,782.3 | 659.1 | 208.8 | 526.6 | 8,676.7 | 79.8 | 38.0 | 80.0 | 280.4 | 239.8 |
| N | 1,668 | 453 | 210 | 692 | 1,603 | 624 | 403 | 197 | 1,156 | 1,455 | 1,612 | 459 | 216 | 684 | 1,706 |

Note: Estimated and actual costs are shown in USD in millions in 2015 prices. Cost overrun is actual cost divided by estimated cost.

[8]For projects that were measured in other currencies (e.g., pound), we converted them to USD 2015 prices.

[9]Costs do not include costs of maintenance, operations, and reinvestments. Estimated cost is the cost that would be incurred to deliver the project based upon the final business case. Estimation practices were found to be similar across projects, with budgets being an outcome of organizational negotiations. Estimates typically include little contingency (up to 15%). Estimates were typically presented as "most likely" cases.

[10]The cost overruns (the variable we investigate in this paper) between ERP and all other project types are statistically significantly different ($p < 0.001$, pairwise Wilcoxon tests with Holm adjustment). The overruns between HRM, MIS, and SCM are not ($p > 0.072$). However, as discussed later in the paper the underlying distributions were not found to be significantly different. The nature of the distribution causes conventional tests (e.g., t-test, Wilcoxon) to be unreliable.

**Table 4.** Descriptive Statistics by Public vs. Private Sector

| | Estimated cost | | Actual cost | | Cost overrun | |
|---|---|---|---|---|---|---|
| | Private | Public | Private | Public | Private | Public |
| Mean | 4.0 | 8.8 | 12.5 | 15.7 | 1.4 | 2.1 |
| Median | 0.4 | 1.7 | 0.2 | 2.2 | 0.9 | 1.0 |
| Sd | 15.8 | 18.9 | 255.7 | 115.2 | 8.2 | 8.9 |
| Min | 0.0005 | 0.001 | 0.0002 | 0.001 | 0.003 | 0.0014 |
| Max | 215.9 | 593.5 | 8,676.7 | 4,376.2 | 280.4 | 239.8 |
| N | 1,755 | 2,691 | 1,156 | 2,679 | 1,748 | 2,759 |

Notes:
1. statistics shown in this table do not include projects where sector is unknown
2. Estimated and actual costs are shown in USD in millions in 2015 prices. Cost overrun is actual cost divided by estimated cost.

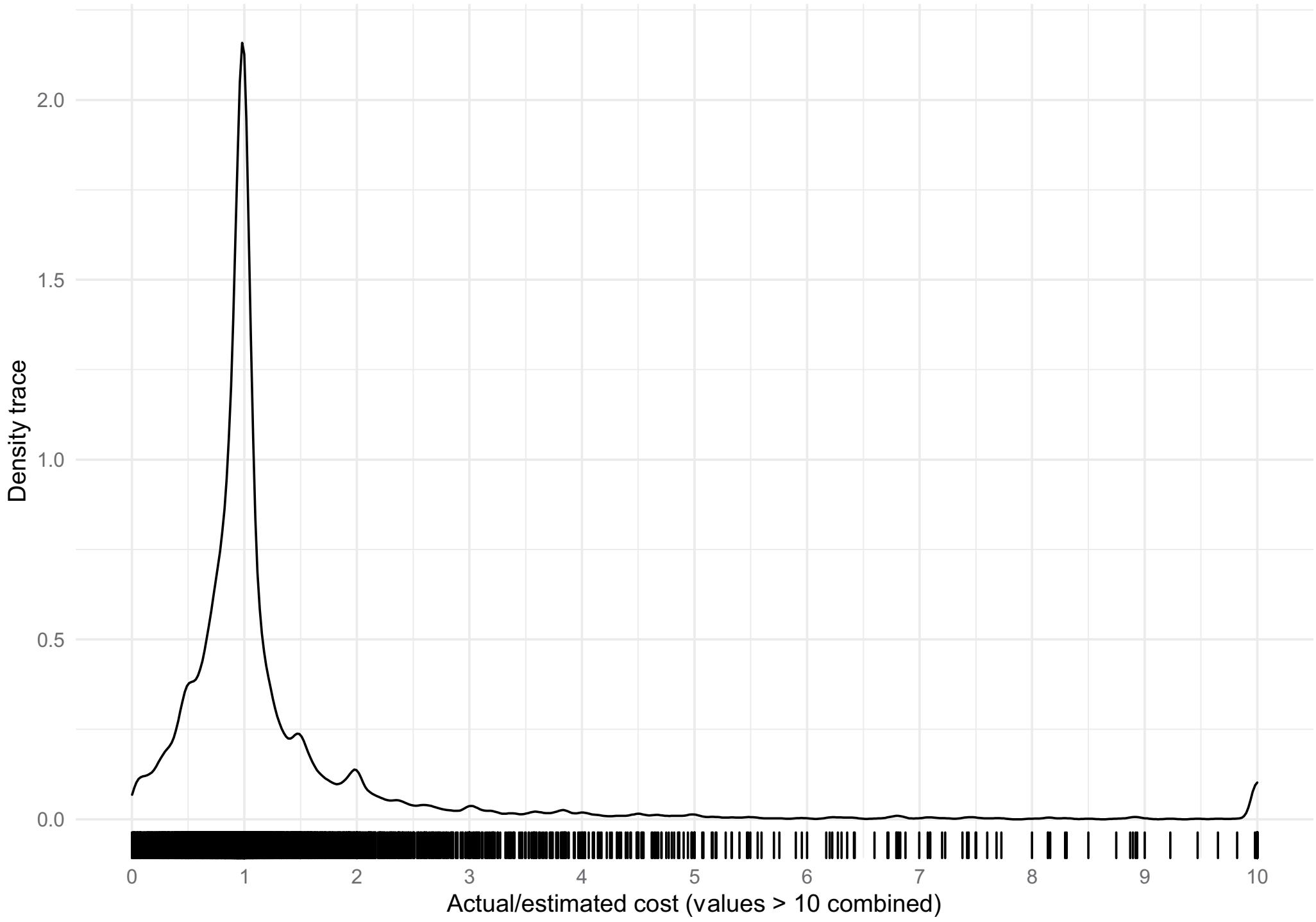

**Figure 1.** Probability Distribution of Cost Overruns (actual cost divided by estimated cost) for 4,677 IT Projects. Values over 10 are combined, which explains the bump at the far right of the curve.

## *Fitting a Power-Law Distribution*

After visual inspection[11] of the probability distribution of cost overruns, we fitted the data to a power-law distribution. The power-law distribution is defined as the probability density function:

[11]We inspected the QQ plots of a range of distributions (Normal, Student's t, Lognormal, Weibull, Beta, Gamma, Exponential, Generalized Extreme Value, Weibull) fitted to the empirical data. A probable distribution is characterized by approximating a straight, 45-degree line in the QQ plots (see Appendix C). In this case, the QQ plots suggest that the power-law distribution is a good candidate for fitting and that the lognormal might be an alternative candidate.

$$f(x; x_0, \alpha) = \frac{\alpha - 1}{x_0}\left(\frac{x}{x_0}\right)^{-\alpha}, x > x_0, \alpha > 1\,. \tag{1}$$

The probability distribution has two parameters: $x_0$ and α. First, $x_0$ is the lower bound of power-law behavior. Second, α is the scaling parameter, which describes the fatness of the tail; a lower α indicates a fatter tail. The scaling parameter α depends on the lower bound of the power-law behavior $x_0$. For fitting our data to a power-law distribution, we followed the procedure recommended by Clauset et al. [19] which has been commonly used for empirically determining whether data can be described as a power-law distribution (e.g., Aguinis et al. [2]). This procedure involves the following steps: (1) estimating the scaling parameter α, (2) estimating the lower bound of power-law behavior $x_0$, (3) examining uncertainty in the parameter estimates for $x_0$ and α using a bootstrapping technique, (4) testing the power-law hypothesis, and (5) comparing the power-law model to an alternative fat-tail model. These steps together help researchers present strong and converging evidence indicating the presence of a power-law distribution in the data [19].

In the first step, we set $x_0$ to overruns above 0% and estimated the scaling parameter using the Maximum Likelihood estimator (see Appendix B in Clauset et al.[19] for the derivation). The initial maximum-likelihood estimate of α for all $x > x_0 = 1.0$ was 3.2.

In the second step, a parameter scan is conducted for each possible value of $x_0$ to find the minimum Kolmogorov-Smirnov (KS) statistic which measures a distance between the empirical distribution function of the sample and the theoretical distribution of the reference distribution (in our case the power-law distribution). Through this procedure, we found $x_0 = 2.0$. We then re-estimated α for all $x > x_0 = 2.0$, which led to an α of 2.3.

In the third step, we used a bootstrapping procedure (with 5,000 iterations) to estimate the uncertainty in the parameter estimates ($x_0$ and α). Through this procedure, we found that $x_0$ has a cumulative mean of 2.05 and a cumulative standard deviation of 1.22, and that α has a cumulative mean of 2.35 and a cumulative standard deviation of 0.20 (see Figure 2). The number of observations in the tail, i.e., $x > x_0$ was on average 791, with a cumulative standard deviation of 485.

In the fourth step, we conducted a test to examine whether our data actually follows a power-law distribution. Clauset et al. [19] suggest a goodness-of-fit test, using a bootstrapping procedure, where the null hypothesis ($H_0$) states that the data is not different from a power-law distribution. We conducted two separate analyses, one using the Anderson-Darling statistic and one using the Kolmogorov-Smirnov statistic. The two analyses yielded p-values of 0.18 (Anderson-Darling) and 0.27 (Kolmogorov-Smirnov), suggesting that our data is not significantly different from a power-law distribution.

In the final step, we compared the power-law fit to an alternative distribution using a likelihood-ratio test. We chose the lognormal distribution as an alternative fit as it is the alternative distribution most commonly compared to and tested against a power-law distribution. We used the exact same five steps to fit our data to a lognormal distribution. First, the estimates for the lognormal distribution are log-mean μ = 0.56 and scale σ = 0.72 for all $x > x_0 = 1.0$. Second, we found the lower bound of $x_0 = 10.0$, and the tail was fitted to all $x > x_0 = 10.0$. This led to new estimates: μ= 1.32 and σ = 1.76. Third, the bootstrapping procedure was used to assess the uncertainty of the parameters with 1,000 iterations. $x_0$ had a cumulative mean of 7.28 with a cumulative standard deviation of 3.19. μ had a cumulative mean of -11.95 with a cumulative standard deviation of 17.18. σ had a cumulative mean of 3.42 with a cumulative standard deviation of 1.97. The number of observations in the tail, i.e. $x > x_0$ was on average 102.19, with a cumulative standard deviation of 54.54. Next, we conducted two separate analyses, one using the Anderson-Darling statistic and one using the Kolmogorov-Smirnov statistic. The two analyses yielded p-values of 0.02 (Anderson-Darling) and 0.03 (Kolmogorov-Smirnov), suggesting that our data is significantly different from a lognormal distribution.

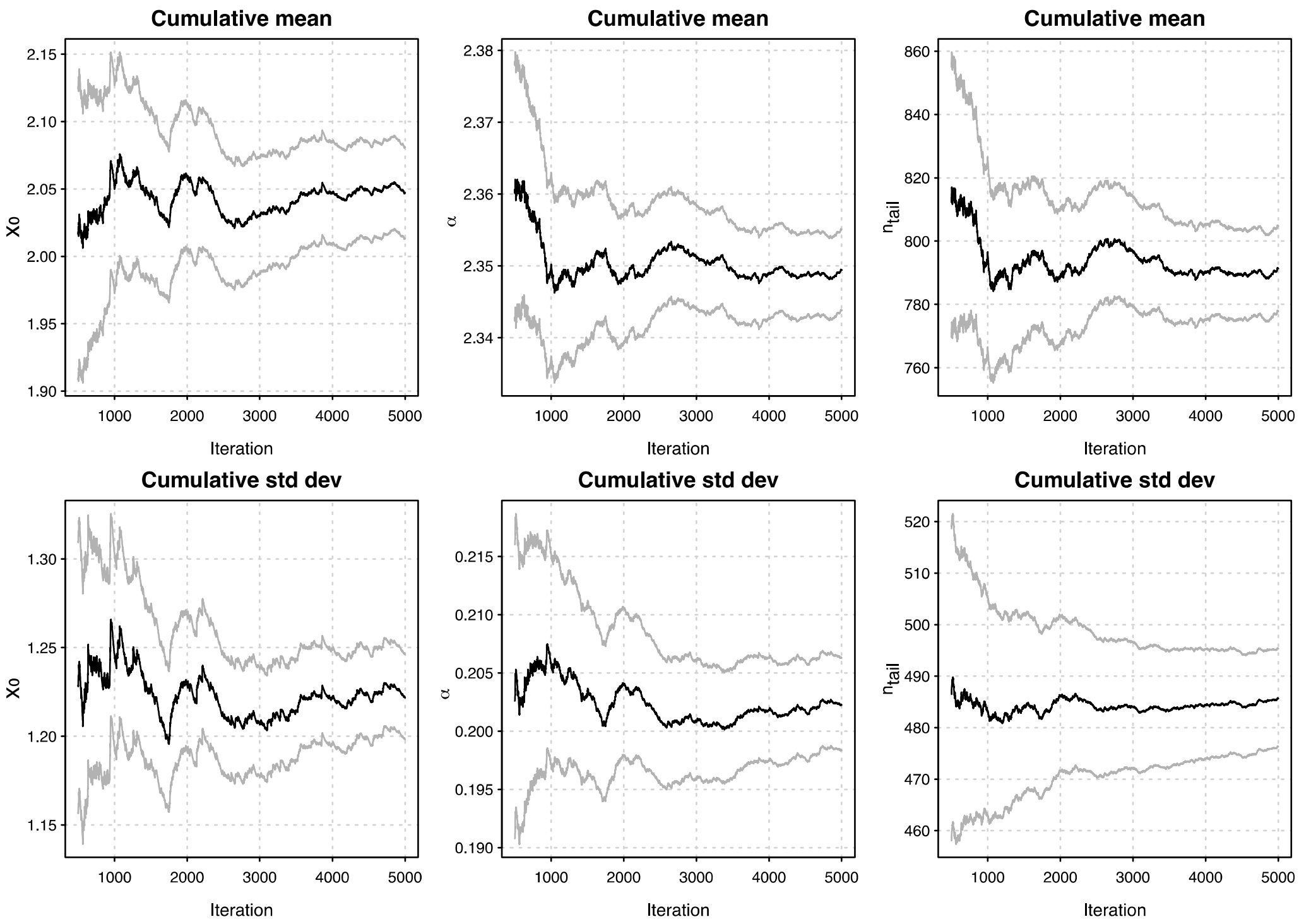

**Figure 2.** The Uncertainty in the Parameter Estimates ($x_0$, α, $n_{tail}$) The first column (left) shows the bootstrapping results for $x_0$.The second column (middle) the boostrapping results for the scaling parameter α. The third column (right) shows the bootstrapping results for the number of observations in the tail, where $x > x_0$.The top row shows the cumulative mean and the bottom row shows the cumulative standard deviation of the estimates.

Finally, we compared the power-law fit with the lognormal fit using Vuong's test [73]. Vuong's test compares the likelihood of either fit in the presence of empirical data. To compare the power-law and lognormal fits the minimum value needs to be set to the same cut-off. For all $x > x_0 = 1.0$ the ratio of the log-likelihood of the two fits was 3.55 in favor of the power-law fit ($p < 0.001$). These results indicate that the power-law distribution is clearly a better fit to the data than the lognormal distribution. Taken together, our results offer strong evidence that IT project cost overruns follow a power-law distribution.

## *Fitting the Tail*

From the analysis above, we found that our data of IT project cost overruns is best described by the following power-law distribution:

$$p(x|\alpha) = 0.65\left(\frac{x}{2.0}\right)^{-2.3}, \qquad \text{for } x > 2.0. \tag{2}$$

For $x_0 = 2.0$ we have 26.5% of the observations in the tail (495 observations). We observe that with an α of 2.3 (sd = 0.058) for the best fit resulting from this approach we find $2 < \alpha < 3$, which entails that the first moment is finite and therefore the mean can be estimated, whereas the second moment cannot, which indicates that the variance is undefined.

The power-law distribution with an α of 2.3 fits the largest number of observations in the right side of the distribution, shown by the dashed curve in Figure 3. However, examining Figure 3 suggests that observations with cost overrun greater than 10 fall above the power law, which indicates that this part of the upper tail is fatter than the power-law fit. This pattern is common for the extremes of the upper tail in power-law distributed data [68]. We carried out further robustness checks and found additional evidence supporting the power-law distribution (detailed procedure and results are reported in Appendix B).

Finally, we tested the sensitivity of the estimated α for different cut-off points, $x_0$, between 1 and 45. Figure 4 shows that α tends to become smaller as $x_0$ becomes larger. α becomes smaller than 2 for $7.1 \leq x_0 \leq 18.8$. We see that irrespective of the cut-off point the observations in our dataset may be described by a power law. We further see that the majority of observations can be described by a power law with $2 < \alpha < 3$, but that parts of the upper tail follow a power law with $\alpha < 2$, indicating extreme fatness. These are the observations that really matter in terms of extreme risk.

## *Ruling Out Rival Explanations*

In this section, we describe additional steps that were taken to rule out rival explanations and to address issues that may have influenced our findings. One rival explanation that has been leveled against observations of power-law tails is that they may be the spurious result of mixing normal distributions with the same mean but different variances [28]. This rival explanation represents a potential issue by suggesting that the power-law distribution observed in our data could be the result of having multiple sizes or types of IT projects in the data. One could argue, for example, that projects with a smaller estimated cost might be subject to a less stringent estimation and evaluation process, with less attention from top management. Thus, while such projects might have a similar mean with larger projects in terms of cost overrun, the less stringent estimation and evaluation process might lead to an increased variance in cost overruns for smaller projects. To rule out this rival explanation we split the dataset by project size (Figure 5). We found that the complementary cumulative distributions for different project sizes are similar and, most importantly, all show the characteristic power-law tails (see Figure 5). The differences between the power-law tails were not found to be statistically significant ($p = 0.863$, Vuong's test, adjusted for multiple comparisons). In addition, we also examined the distribution of IT project estimated costs and found that our data includes a large range of project sizes without any bias towards smaller or larger projects.[12]

[12]See Appendix D for these results as well as additional information and analyses on the distribution of actual project costs and the distribution of cost overruns in absolute value (the overspend expressed in USD millions).

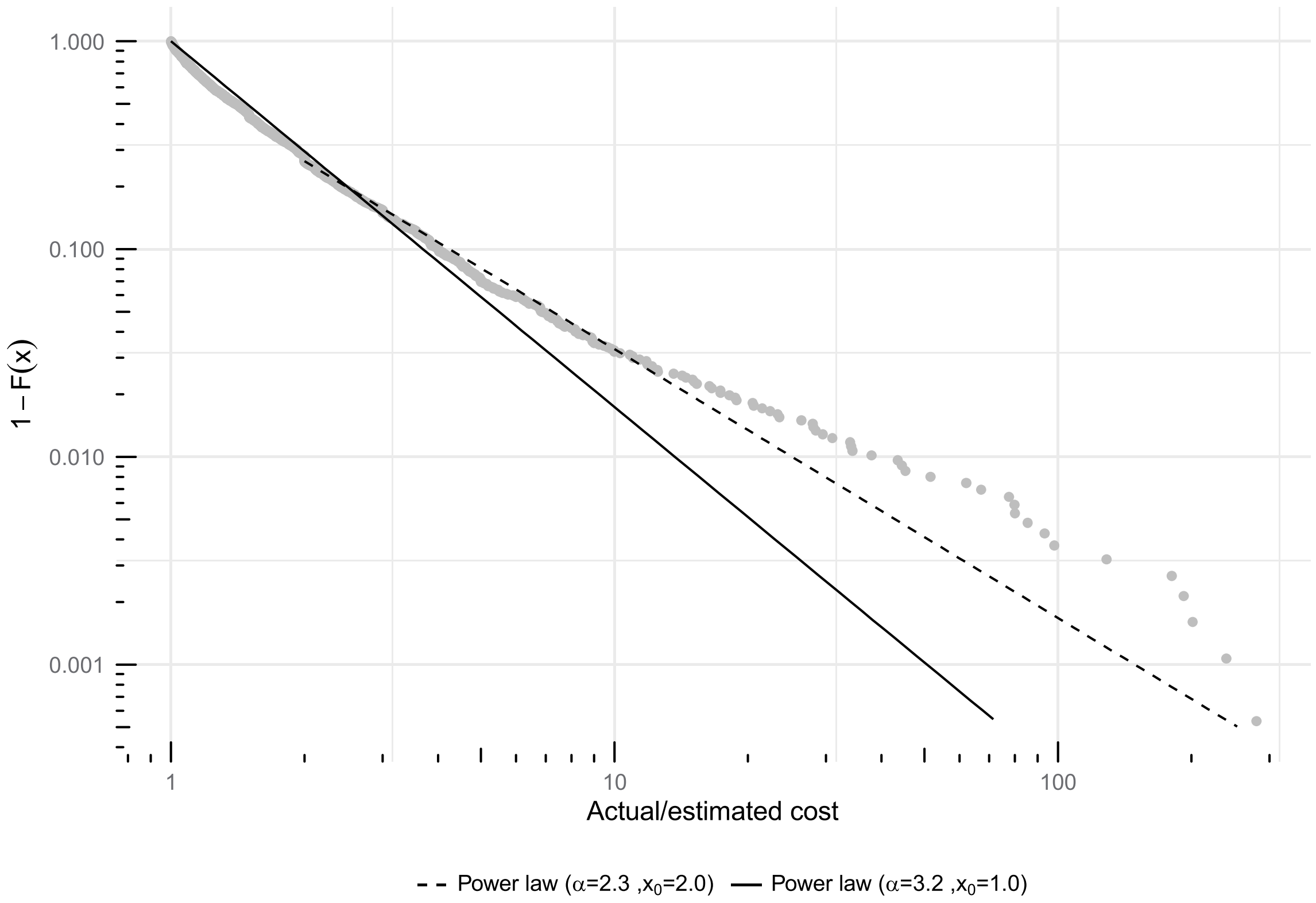

**Figure 3.** Fit of Power-Law Distributions to the Upper Tail of Cost Overruns x (horizontal axis) is actual cost divided by estimated cost. F(x) is the cumulative distribution function, with 1-F(x) (vertical axis) showing the probability of cost overrun being larger than x (log-log plot). The two power-law distributions are shown.

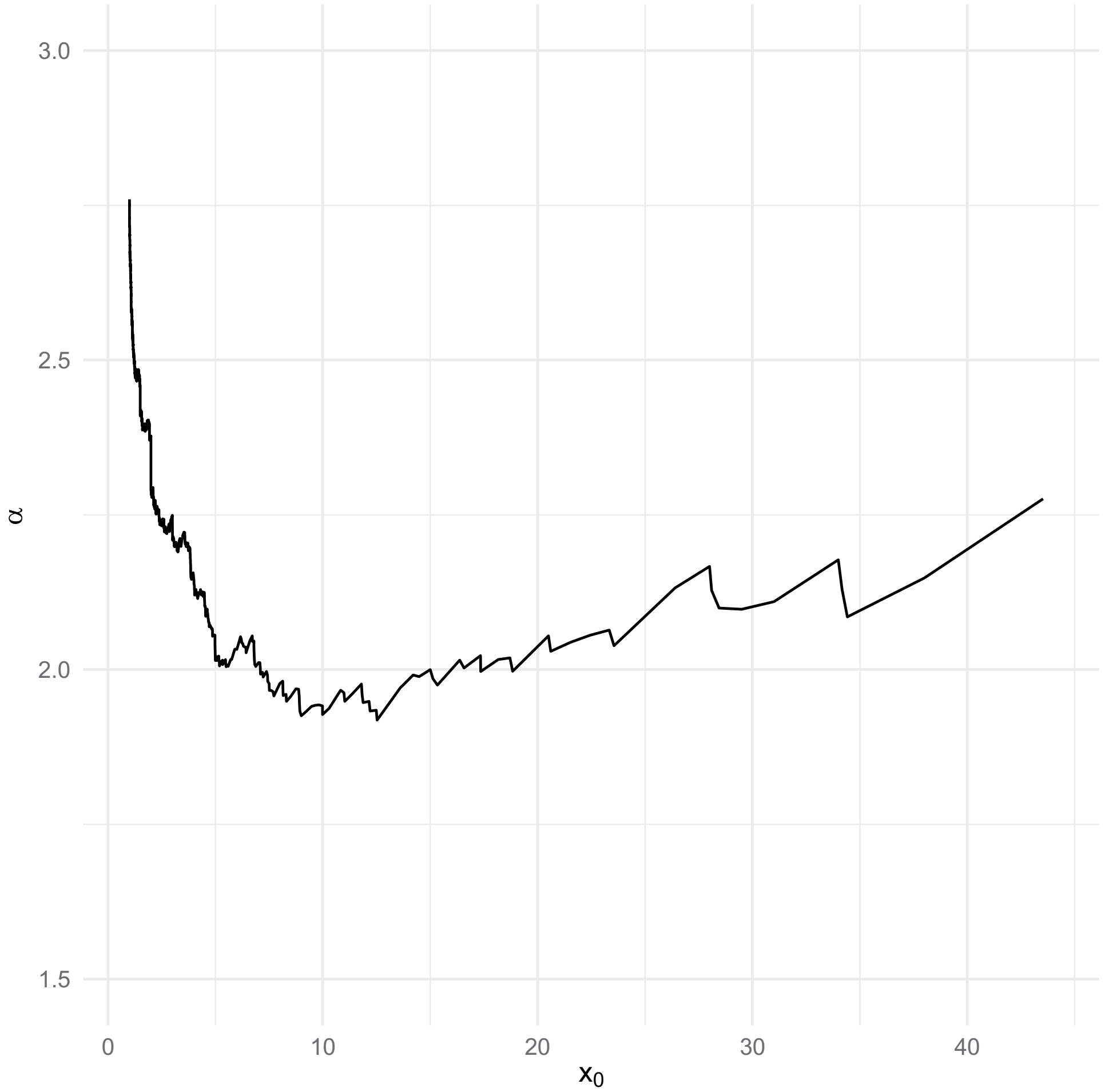

**Figure 4.** Estimated power-law fit for $1 < x_0 < 45$ $x_0$ (horizontal axis) is the lower bound of the power-law behavior. α is the mean estimate of the fitted scaling parameter of the power law.

The projects in our data also varied by type. Most frequent are supply chain management (25%), enterprise resource planning (19%), and human resource management (9%). Since it is possible that the power law observed in our data is a result of mixing normal distributions for different project types with similar means but different variances, we examined this as well. Figure 6 shows the cumulative distributions of the cost overruns by project type. The data again show the pattern of power-law tails with no statistically significant differences of the distribution between project types ($p > 0.092$, Vuong's test with adjustment for multiple comparisons).

Since we collected data on IT projects from two different sources, we also checked to see if there were any statistically significant differences. We found no statistically significantly differences ($p = 0.648$) between the data from the two sources, i.e., U.S. federal projects and projects obtained from participating firms. Similarly, we did not find any statistically significant differences in cost overruns ($p = 0.933$) between projects of short, medium, and long project duration. Finally, we explored whether there was a systematic bias between single projects (i.e., those for which there is a one-to-one mapping between project and organization in our dataset) and multiple projects (i.e., those for which there is a many-to-one mapping between project and organization in our dataset). We found that both the single projects (n=593) and the multiple projects (n=4,084) followed a power-law distribution and that there was no statistically significant difference in the power-law fits between the two groups of projects (Vuong's test p = 0.492). By examining projects of different sizes, types, data sources, project durations, and comparing single projects and multiple projects we were able to rule out a number of rival explanations and show that there is robustness in our observed power-law findings.

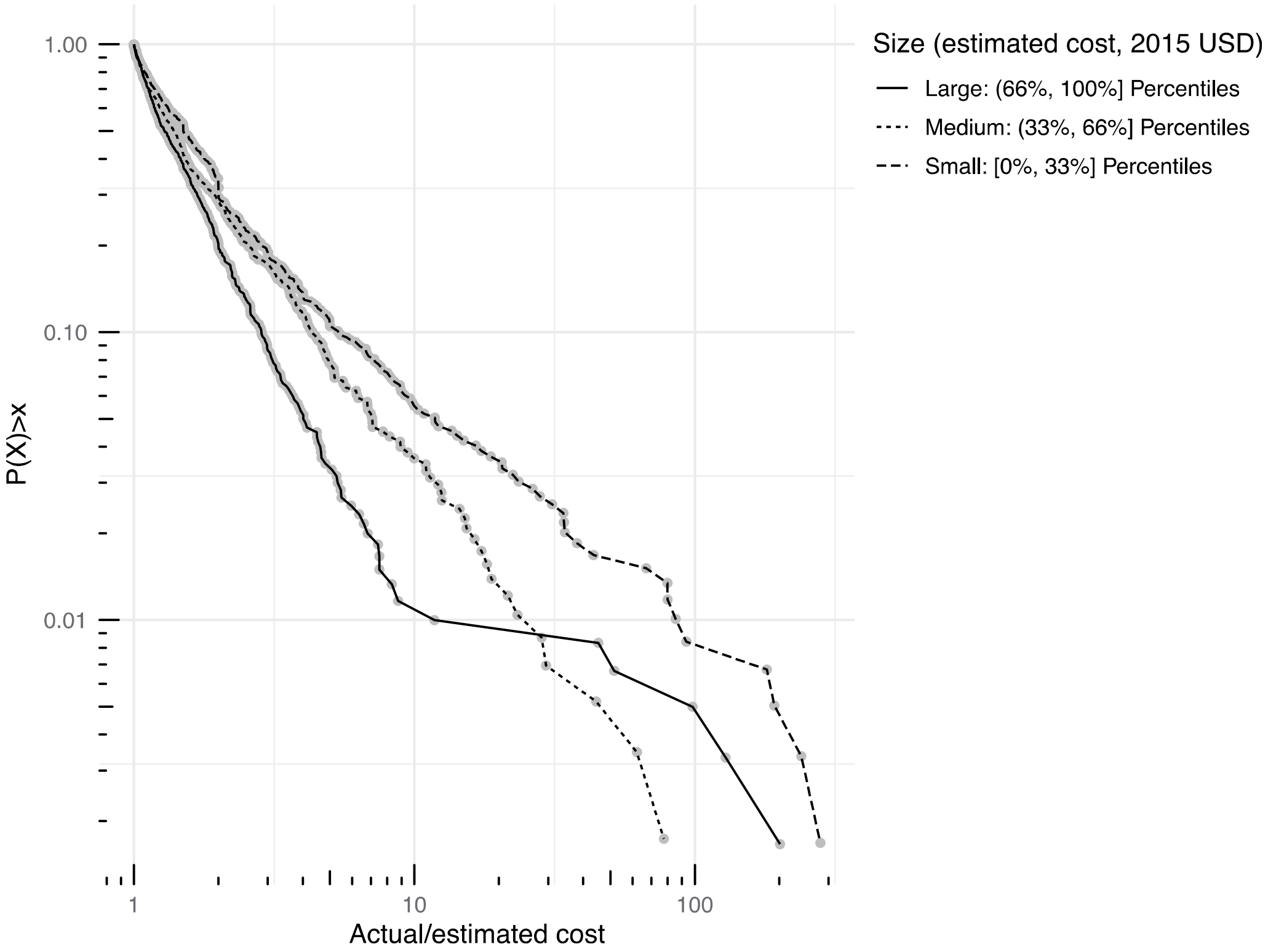

**Figure 5.** Complementary Cumulative Distribution of Cost Overruns (expressed as ratio actual/estimated cost) Grouped by Estimated Project Size, Measured in 2015 USD Terms. The three groups are balanced in size by defining group one as all project sizes up to and including the 33rd percentile; group two as all project sizes larger than 33rd percentile, up to and including the 66th percentile; and group three as all projects above the 66th percentile. The complementary cumulative distributions for the three groups are similar and all show characteristic power-law tails.

### *Cost, Effort, and Schedule*

Besides cost, we were able to obtain estimated and actual effort (n = 158) and schedule (n = 962) for some of the projects in our dataset. This allowed us to conduct a power-law fitting of the tail for effort overruns and schedule overruns (see Table 5). In fitting the power law, we followed Clauset's approach as we did before for cost overruns. In addition, for both effort and schedule overruns, we used the Kolmogorov-Smirnov statistic to test whether our data followed a power-law distribution. The KS statistics provided a p-value greater than 0.05 for both effort and schedule overruns and thus we cannot reject the null hypothesis that the data is generated from a power-law distribution). Thus, it appears that in addition to cost overruns, effort and schedule overruns also follow a power-law distribution.

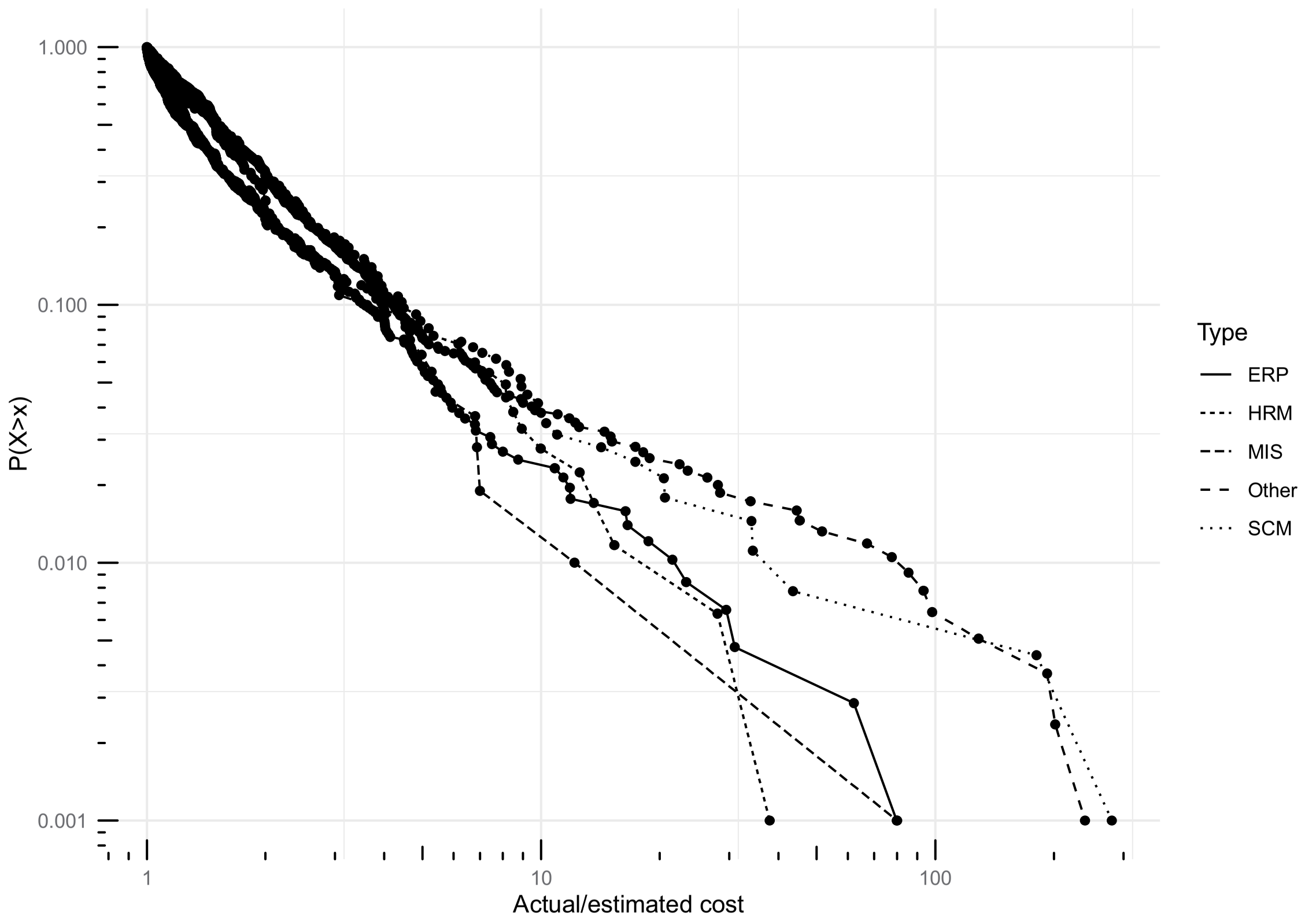

**Figure 6** Complementary Cumulative Distribution of Cost Overruns (actual/estimated cost) Grouped by Project Type. Shown are the most frequent project types in the dataset: supply chain management (SCM), enterprise resource planning (ERP), human resource management (HRM), and management information systems (MIS). The complementary cumulative distributions show the characteristic pattern of power-law tails.

**Table 5.** Results for Power-Law Fitting to Cost, Effort, and Schedule Overruns

| Variable | N | Distribution with best fit | α-value | $x_0$-value |
|---|---|---|---|---|
| Cost overrun | 4,677 | Power law | 2.3 | 2.00 |
| Effort overrun | 158 | Power law | 2.6 | 1.49 |
| Schedule overrun | 962 | Power law | 3.3 | 1.08 |

## Generative Mechanism and Computer Simulation

### *Theoretical Background*

In this section, we theorize how interdependencies among technological components [8, 21, 72] in an IT system can contribute to generating a power-law distribution of IT project cost overruns. Interdependencies produce certain behavioral patterns in systems; for example, "if something in Component 1 changes, then Component 2 may need to change as well" [21, p. 711]. In other words, changes or problems in a technological component can have negative impacts on other interdependent technological components.

The notion of interdependency in IT systems is analogous to interdependency among components in complex systems that are often found in the natural sciences. In a natural forest, for example, clusters of trees are adjacent to one another (interdependency), and a certain event (e.g., a lightning strike) affects not only a single cluster of trees, but other clusters that are connected to the cluster initially affected by the lighting strike [49]. Self-organized criticality suggests that a complex system organizes itself into a critical state in which a single event can trigger a series of cascading events to produce an extreme outcome. Interdependency is a central concept in self-organized criticality, which explains that a small event initially affecting a single component can cause a snowballing effect by affecting other components that are interdependent with the initially affected component.

From the perspective of self-organized criticality, extreme values that we observe in power-law distributions are generated due to interactions among a large number of components in a complex system, and they tend to follow a nonlinear-effects pattern in which the change of the output is not proportional to the change of the input [15]. In other words, an initially affected component can in turn affect not just one, but many other interconnected components. Similarly, catastrophic failures in large IT systems tend to occur when multiple failures occur as a result of interdependencies and those failures compound one another as opposed to when a single, independent component fails [61]. The key point that can be drawn here is that substantial overruns in IT projects may result from interdependencies. Further, substantial overruns may be difficult to predict from the behavior of individual components, and can be better understood through the lens of self-organizing behavior. These lines of thinking suggest that a substantial cost overrun can result when one troubled component of an IT project has interdependencies with other components, creating a cascading situation.

Self-organized criticality has been discussed by prior research as a generative mechanism for power-law distributions observed in organizational contexts [3, 16] as well as physics [58]. For example, in organizational contexts the self-organized criticality perspective can help shed light on why a small number of individuals produce extremely large outputs far superior to what most people produce. Researchers suggest that top scientists tend to focus on developing a research program where multiple and related projects are being undertaken simultaneously [66]. In this setting, a single breakthrough on one particular project can lead to a number of breakthroughs on other projects that are related. Further, the magnitude of the total breakthroughs is determined by the total number of projects and how intricately those projects are related with one another.

In physics, the forest-fire model has been developed and commonly used to demonstrate how self-organized criticality contributes to producing power-law distributions that describe the magnitude of forest fires. The forest-fire model, when used in computer simulations, has been shown to produce data that is generally consistent with data observed in actual forest fires [49]. The model begins with an assumption that the forest can be viewed as a square grid consisting of a number of tree clusters. Each cluster in the grid can be empty or occupied by trees. Each cluster is neighbored by four adjacent clusters. Lightning strikes a cluster at random and this starts a fire in the cluster. Any cluster filled with trees that lies adjacent to the cluster struck by the lightning also catches fire. Interconnectedness represented by adjacency of clusters and the randomness of a lightning strike represent two of the defining features of the forest-fire model. The total number of clusters destroyed by the fire (the magnitude of the forest fire) is determined by how intricately the tree clusters are connected with one another [58].

The forest-fire model, we argue, not only illustrates the concept of self-organized criticality and the role that interconnectedness plays in producing power-law distributions, but can also be extended to help theorize why and how IT project cost overruns might follow a power-law distribution. Specifically, randomness in selecting dependencies among components in an IT system is analogous to randomness in how intricately tree clusters are connected. Tree clusters become connected with one another not by design but, rather, organically [58]. In an organization, no matter how well planned an IT system is, some of the interdependencies between technological components cannot be anticipated because they occur organically as the system is developed and as the development team responds to changing requirements. This aspect of systems development can be understood as a condition that leads to self-organized criticality. Furthermore, in building an IT system any cost overrun in one component is expected to lead to cost overruns in other interdependent components, and this overrun event (and chain reaction) is analogous to the lighting strike causing a fire in a tree cluster subsequently affecting other adjacent clusters in the forest-fire model.

## Simulation Design

Drawing upon the notion of interdependencies as an essential characteristic of IT systems and the idea of self-organized criticality which provides the conceptual foundation for the forest-fire model, we created a computer simulation to examine the power-law distribution of IT project cost overruns. We aimed to create a simulation for an IT project that would be similar in size to building an IT system for a business division of a large company and used an empirical finding by Mocker [55] as a benchmark. Specifically, Mocker [55] identified 273 applications being used to support various business processes in the investment banking division of an European bank. Therefore, we created 300 technological components underlying a new IT system project.[13]

Next, we created interdependencies among technological components. Our selection of parameters for interdependencies was informed by prior research. Specifically, prior research found that the degree of interdependencies does not follow a normal distribution and outlier/extreme values exist and should not be removed from the data [42, 55]. For example, Khosroshahi et al. [42] found that extreme values are more than 10 times larger than the mean of 95% of the data. Similarly, Mocker [55] found that there is a significant difference between the mean and the 95% trimmed mean for IT system dependency data. Therefore, in our simulation we seeded a large number of components with no or low interdependencies and two components with extremely high interdependencies. Specifically, we designed our simulation so that 148 components (roughly 50% of the 300 components) did not have any dependency. Further, we designed the simulation to randomly select one component (out of 300) to have a dependency with 75 other components and another component (out of 300) to have a dependency with 25 other components. The extremely high interdependencies associated with these two components is similar to that which has been found in prior research [42, 55]. Further, the simulation randomly selected 50 (out of 300) components to have a dependency with one other component, another 40 (out of 300) components to have a dependency with two other components, another 30 (out of 300) components to have a dependency with three other components, another 20 (out of 300) components to have a dependency with four other components, and another 10 (out of 300) components to have a dependency with five other components.

Next, for simulation purposes, we assumed it would take one cost-unit to complete each component. However, we designed the simulation to randomly select one component to experience a cost overrun. Further, this overrun could range from costing anywhere between 0.1 and 280 additional cost-units (random selection was based on the uniform distribution).[14] We chose this range because in our empirical data an overrun of 280 (ratio of actual/estimated costs) was the maximum value observed. In terms of the way in which cost overruns propagate in our simulation, the first component which experiences an overrun affects other components connected to first component (we assumed, for the sake of simplicity, that these overruns would be of the same magnitude[15]), but the components affected by the first component do not affect the first component (i.e., there is no recursion in the model). In addition, subsequent components affected by the first

[13]We note that prior research using the forest-fire model employed a much larger system with a greater number of components [49]. However, we chose a number of components that would be more realistic in the IT project context.

[14]We believe this was a more conservative approach because a power-law distribution of cost overruns in individual components is more likely to lead to a power-law distribution of cost overruns at the project level. In addition, as demonstrated later in the results section, while overruns in individual components follow a uniform distribution, it is found that overruns at the project level end up following a power-law distribution. This suggests that the pattern of project overruns is more than just a sum of overruns in individual components, and that the dependencies of individual components may be key to understanding the power-law of cost overruns in IT projects.

[15]This was a simplifying assumption that we made for two reasons: (1) there is no guiding theory or empirical knowledge that would inform us as to what magnitude of knock-on effects might occur when one component experiences an overrun, and (2) we wanted to keep the model tractable. In future research, it may make sense to further refine the model.

**Table 6.** Procedure of Computer Simulation

| |
|---|
| Step 1. Create 300 technological components |
| Step 2. Randomly create connections between components (see note 1 below for detail) |
| Step 3. Assign cost ($x_{max} \sim n^{1/(\alpha-1)}$) associated with each component |
| Step 4. Randomly select a component to experience an overrun |
| Step 5. Randomly determine the degree of overrun (X=$C_i = 1$) |
| Step 6. Overrun is cascaded to other components connected to the initial component selected in Step 4 (see note 2 and 3 below for details) |
| Step 7. Update cost (1.1~280) of components (see note 4 for detail) |
| Step 8. Calculate total costs (sum of costs of all components): $TC = \sum_i^{300} C_i$ |
| Step 9. Steps 1-8 are repeated 5,000 times |

Note 1:
50 out of 300 components have a connection with 1 other component
40 out of 300 components have a connection with 2 other components
30 out of 300 components have a connection with 3 other components
20 out of 300 components have a connection with 4 other components
10 out of 300 components have a connection with 5 other components
1 out of 300 components has a connection with 25 other components
1 out of 300 components has a connection with 75 other components
148 out of 300 components have no connection with other components
Note 2:
This process is non-recursive in the sense that the first component affects other components connected to the first component, but the components affected by the first component do not affect the first component.
Note 3:
This process is iterative in the sense that subsequent components affected by the first component can affect other components that are connected to them. This process continues until subsequent components have no additional connection with other components.
Note 4:

$$C_i = \begin{cases} 1, & \text{if not affected by overrun} \\ 1 \times X, & \text{if affected by overrun} \end{cases}$$

component can affect other components that are connected to them. This process continues until subsequent components have no additional connection with other components. A summary of our simulation design is shown in Table 6.

Figure 7 illustrates interdependencies among technological components in our simulation (based on a single project). In this network graph, the direction of an arrow indicates an interdependent component that can be affected by another component that experiences an overrun.

## Simulation Data and Results

We ran the simulation 5,000 times, each producing the cost performance of an IT project. With no overrun, each project would consume 300 cost-units. Each simulation produced the total cost units for a single project. We created an overrun ratio (the total cost units/300) for each project, and this data was used for the analysis.

We followed the approach used by Clauset et al. [9] for fitting the simulation data. We first estimated α (fatness of the tail) for all values of $x_0$ (lower bound of the tail) and found that the simulation data can be best fitted to a power-law distribution with the scaling parameter α of 2.5 for all $x > x_0 = 28.7$. Next, we used a bootstrapping procedure (5,000 iterations) to calculate ranges of the parameter estimates (cumulative mean and cumulative standard deviations) (results shown in Figure 8). Through this procedure, we found that $x_0$ has a cumulative mean of 29.6 with a cumulative standard deviation of 27.9. We found that α has a cumulative mean of 2.7 with a cumulative standard deviation of 2.0. The number of observations in the tail, i.e., $x > x_0$ was on average 881, with a cumulative standard deviation of 776. Finally, we used the Kolmogorov-Smirnov statistic to judge if the distribution of the simulation data is significantly different from a power-law distribution. The Kolmogorov-Smirnov statistic was not statistically significant ($p = 0.07$), suggesting that the distribution of the simulation data is not significantly different from a power-law distribution. Based on these results, our computer simulation demonstrates that interdependencies among technological components in an IT system combined with self-organized criticality can contribute to generating the power-law distribution of IT project cost overruns.

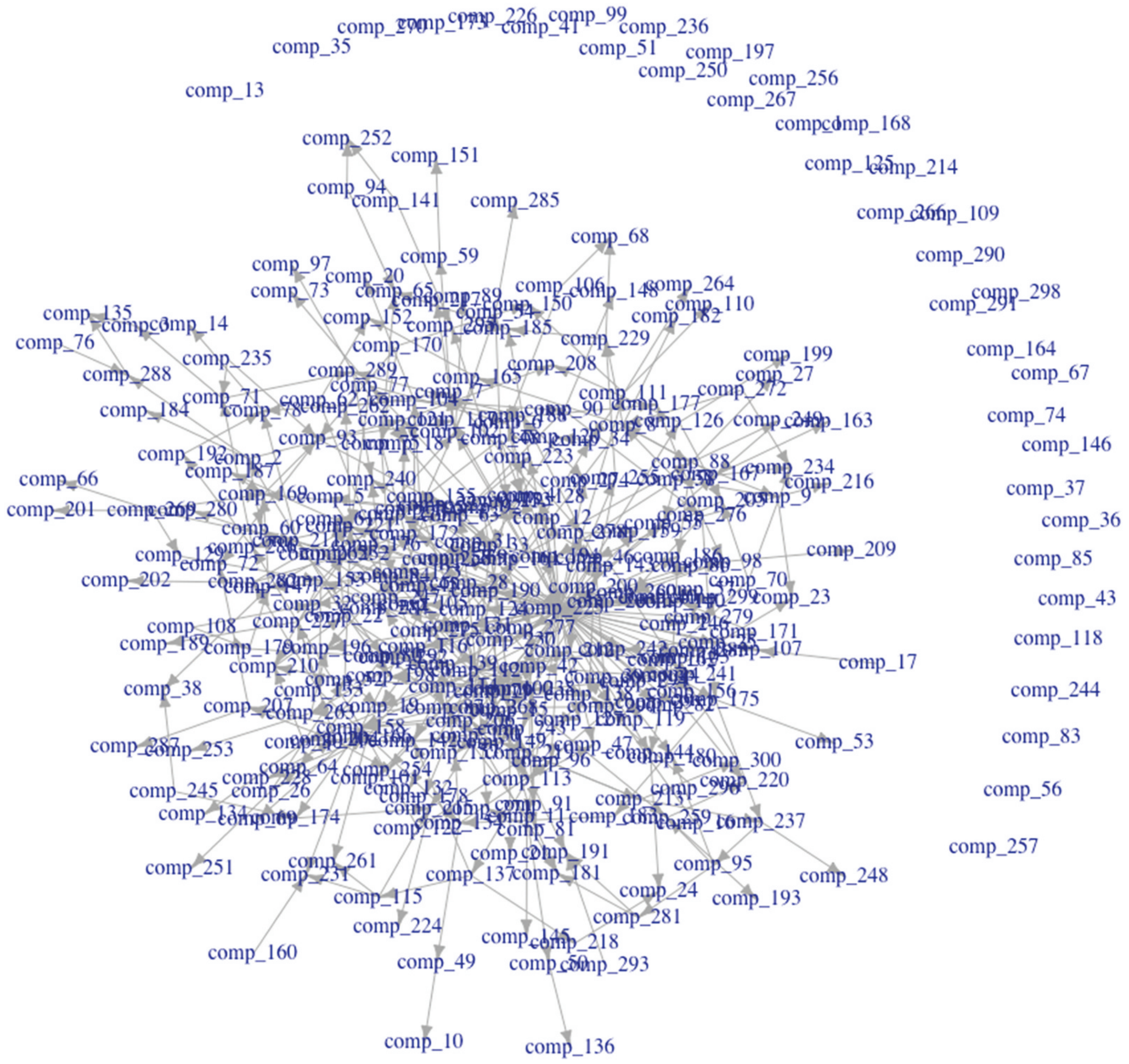

**Figure 7.** An Illustration of Interdependencies Among Technological Components

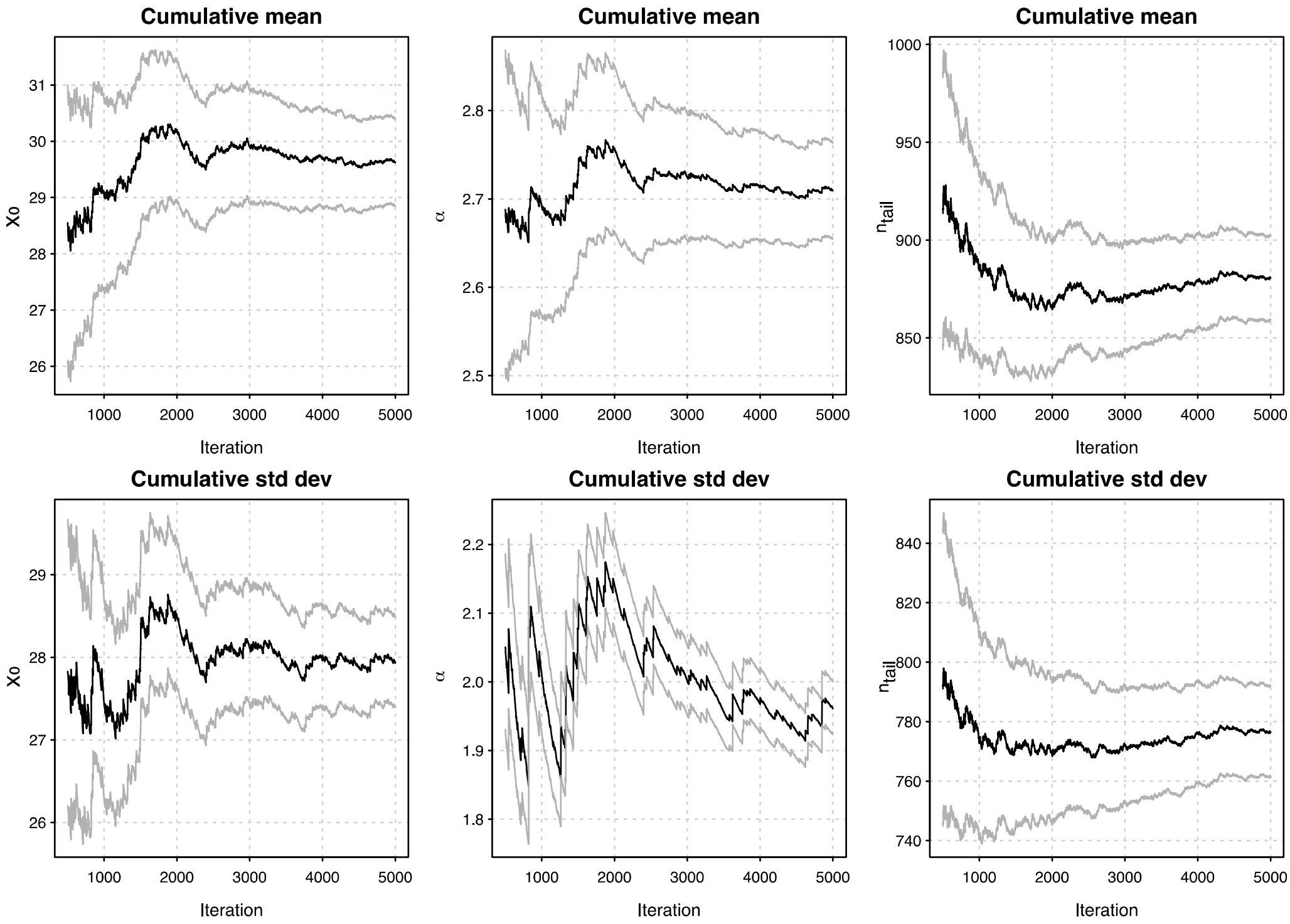

**Figure 8.** The Uncertainty in the Parameter Estimates ($x_0$, α, $n_{tail}$) – Simulation Data The first column (left) shows the bootstrapping results for $x_0$.The second column (middle) the boostrapping results for the scaling parameter α. The third column (right) shows the bootstrapping results for the number of observations in the tail, where $x > x_0$.The top row shows the cumulative mean and the bottom row shows the cumulative standard deviation of the estimates.

## General Discussion

In this research, we assembled a large dataset of IT projects with a focus on actual and estimated project costs to shed light on the probability distribution of cost overruns in IT projects (RQ1) and proposed a generative mechanism to explain the probability distribution of cost overruns in IT projects (RQ2). Building a large dataset was particularly important in this research to understand the probability distribution of IT project cost overruns, as extreme events are rare and tend to be underrepresented in small datasets [33, 51]. Our research shows that a certain range in the tail ($7.1 \leq x_0 \leq 18.8$) carries the potential for extreme risk, where variance is infinite. Our findings show that disastrous IT projects are not the outliers they are sometimes interpreted as, but are instead extreme values that follow a highly regular and predictable power-law pattern. For this range in the tail, α is smaller than two, which means the tail is so fat that neither mean nor variance exist. Regression to the mean is a meaningless concept when this occurs, whereas "regression to the tail"[16] is real and consequential [29]. In short, the average cost overrun for IT projects does not exist (i.e., cannot be calculated), [17] and this should be a sobering thought for anyone concerned with the risk associated with financing IT projects. In addition, our research suggests that assuming a normal or near-normal distribution can lead people to significantly underestimate the risk of cost overruns in IT projects (see Appendix E for discussion concerning the magnitude of error that can be caused when a normal or near-normal distribution is assumed). In this section, we discuss the implications of our work for research and practice, as well as limitations and directions for future research.

### *Implications for Research*

First, having discovered that IT project performance fits a power-law distribution and that we are dealing with fat tails when it comes to IT project performance, we suggest that the commonly used measures of mean and standard deviation may not be adequate to represent or fully comprehend IT project performance data. Other measures, such as median, quartile, interquartile range, and median absolute deviation, should be examined and reported in studies focused on IT project performance. Such measures combined with boxplots may reveal that, contrary to conventional wisdom, there is a strong tendency for IT projects to be on budget (mode and median close to 0% overrun).

Further, our review of prior IS research on IT project performance suggests that a majority of studies tend to report only mean and standard deviations of project overruns, and it is unclear how they deal with maximum values. In addition, some studies used a fixed survey scale (e.g., the extent to which the project was completed within budget and schedule on a 5-point likert scale) for measuring project overruns (see Jiang et al. [36] and Keil et al. [40] for example), and we suggest that such an approach may not be appropriate given the power-law nature of IT project overruns discovered by this research. We suggest that using a more precise measure (e.g., a proportion of overrun in comparison to original budget or schedule) is warranted.

Second, to date, IS researchers have tended to treat project performance (e.g., cost overrun) as an outcome variable in a causal model. While causal models are useful for understanding factors influencing project performance, they are not capable of explaining substantial overruns (extreme values) in IT projects. In other words, the existing body of knowledge offers no explanation as to why some IT projects experience substantial overruns as observed in this research. Therefore, this research makes an important contribution to IS research on IT project management by introducing a theoretical explanation based on self-organized criticality [7, 35] and interdependencies. Specifically, this research suggests that a small proportion of IT projects experience extreme cost overruns due to interactions of interconnected components of an IT system. In particular, when a project organizes itself into a critical state, an overrun in a single technological component can trigger cascading reactions involving many other interconnected components. In addition, extreme cost overruns tend to be unpredictable, or nondeterministic [6, 16, 68] because interdependencies are formed based on the nature of IT systems [21].

[16]A distribution must have non-vanishing probability density towards infinity (or minus infinity) for regression to the tail to occur in data sampled from it. This non-vanishing probability density towards infinity looks like a tail on a graph of the distribution. Regression to the tail occurs only for distributions with infinite variance. The frequency of new extremes and the amount by which a new extreme exceeds the previous extreme indicate whether the underlying distribution that data are sampled from is fat-tailed or not, and hence whether or not it has an expected value and finite variance, or infinite variance and hence no well-defined expected value. In the latter case, "regression to the mean" means regression to infinity, i.e., there is no mean value in the conventional sense (the frequentist sense). Ever better attempts to estimate this mean with conventional methods (i.e., by the mean values of a sample) will yield ever larger values, which is to say values in the tail. For simplicity, we disregard the situation with fat tails towards both positive and negative infinity, which would be "regression to two tails," exemplified by the Cauchy distribution.

[17]It should be stressed that the range of the tail in which $\alpha$ is smaller than two, and only this range of the tail, determines this finding. All observations to the left of the tail, including the many observations in the mode, are irrelevant to the outcome and to the existence or non-existence of mean and variance.

Third, our research also adds to the discourse concerning modularity in IT systems. Notions of both interdependencies among technological components (which is an important element of the proposed generative mechanism for the power law) and modularity find their roots in Simon's [65] conceptualization of sub-systems and interactions between sub-systems of large systems. Prior IS research has suggested that segmenting a system's functions into small modules (components) improves design quality and adaptability [76]. In addition, researchers suggest that modularity can be of value when a system is so large that there are numerous elements to the system [27]. Nonetheless, when interdependencies among modules exist this can cause significant delays and errors during implementation, testing and integration phases [27] and in our research interdependencies among technological components were shown to contribute towards generating extreme overruns (fat-fails). Therefore, our research adds to the extant knowledge of modularity in IT systems by suggesting that the presence of interdependencies among numerous components in a large IT system can increase the risk of cost overruns.

Prior research suggests that the complex nature of information systems development contributes to high failure rates [74]. Our research identifies interdependencies among components in an IT system as a potential source of complexity in IS development. Interdependency in an IT system also has implications for research on organization design, particularly with respect to organizing teams for IT projects. People charged with implementing different interconnected components of a large project must communicate with one another and collaborate to manage interdependencies. Prior research suggests that when ties among organizational units mirror interdependencies in the work being performed, it can help reduce coordination costs and improve coordination [8, 21]. In a similar vein, the project team may be able to manage interdependencies among components in an IT system more effectively by aligning ties among people (or teams) with interdependencies among components allocated to them.

Finally, research focusing on IT project performance must aim to gather a large sample size to avoid the pitfall of underweighting rare events in small samples [24]. Small samples may not include projects that experienced significant overruns, and this may lead to a biased conclusion concerning IT project performance. Another implication of this research is that decision makers should not take rare, extreme, overruns lightly. Assuming the normal distribution of decision outcomes, extant normative decision theories recommend that decision makers not shy away from making the choice that resulted in unfavorable outcomes in the past (i.e., don't be afraid of risky and novel choices) [25]. This is because the choice that resulted in unfavorable outcomes in the past is unlikely to continue to produce similar unfavorable outcomes. This wisdom is in fact built upon the concept of regression to the mean. However, such advice can backfire when regression to the tail is present. Therefore, our research offers a way to better understand decision theories built on assumptions of statistical distributions.

## Limitations and Future Research

All research is subject to limitations and ours is no exception. First, while we presented a generative mechanism for IT project cost overrun based on self-organized criticality and interdependencies among technological components in an IT system, the mechanism was not empirically tested. Consistent with prior research that has been done involving self-organized criticality (e.g., forest fire models), we focused on comparing real-world data and simulation data as a proof for the generative mechanism. Such an approach, however, has inherent limitations.

Second, while our research involved assembling a very large dataset of IT projects, it will be useful to see if other similar sized or even larger datasets of IT projects can be assembled and if these corroborate our findings. Further, for our data collection we relied on firms and U.S. federal government agencies that keep track of project performance (e.g., project documentation, audited data, etc.). We note that such organizations may have more mature project management practices than organizations that do not keep track of project performance. Therefore, project performance in our sample may be better than project performance generally.

Third, in this research we do not have data on how overruns occurred over different time periods of a project's lifecycle (e.g., whether overruns tend to occur earlier or later in a project's lifecycle). Therefore, one direction for future research would be to develop a more granular understanding of how cost overruns occur over time.

Fourth, research suggests that decision makers tend to avoid choices that produce greater variability in results even if such choices eventually produce better outcomes in the long run [25]. This suggests that the distribution of decision outcomes can influence decision makers' choice. The findings of this research challenge the common assumption that IT project performance follows a normal (or near normal) distribution. One fruitful direction for future research is to empirically investigate how IT project managers make project-related decisions (e.g., project estimation, risk management choices, etc.) under the condition of normally distributed IT project performance data vs. the condition of non-normally distributed IT project performance data (e.g., power law).

Fifth, in our simulation no intervention that can mitigate cost overruns was considered. Therefore, one might expect the results of the simulation to be worse than what was observed in the empirical data (where interventions were possible). However, this is not what we found. Comparable results between simulation and empirical data indicate that interventions may not have substantial impacts on containing cost overruns and this could be due to the nature of IT-based systems and dependencies among technological components. In the IS literature, there exists a large body of knowledge on mitigating risks in IT projects [10, 75]. Further research is warranted on how and what risk mitigation strategies can be effective in dealing with interdependencies in an IT system and preventing substantial project overruns.

Finally, our data does not allow us to distinguish between agile projects and those that were managed using waterfall approaches. Given the increasing popularity of agile development and its known benefits [5], further empirical research is warranted to compare the performance of agile IT projects vs. non-agile IT projects, and their respective probability distributions.

### Implications for Practice

The implications of our findings for practice are both clear and disturbing. One important consequence of fat tails and the power-law nature of IT projects is that the average expected maximum cost overrun increases exponentially with the sample size, generally with $\langle x_{\max} \rangle \sim n^{1/(\alpha-1)}$, or in the case of IT projects $x_{max} \sim n^{0.77}$. Thus, the average maximum cost overrun in principle has no upper bound, which is an unsettling result for IT project managers and sponsors. The longer the time horizon of observation or the bigger the sample, the more extreme values are expected to be found. The implication is that small samples, like those found in existing studies of IT project performance, will lead to bias that underrepresents the real risk of cost overruns.

Given the fat tailed distribution identified for IT project cost overruns, extreme cost overruns are to be expected. In fact, our research suggests that no matter what their prior experience has been, decision makers should not assume that there will not be another, even larger cost overrun, in their future. The power-law nature of IT performance explains the anecdotal evidence in the introduction, with careers and companies being destroyed by extreme cost overruns. These disastrous outcomes are a consequence of the power law discovered above. What the power law of IT project cost overruns tells us is that extreme IT project cost overruns will occur. We cannot predict when the next one will happen or how big it will be, but we *can* predict that (1) more will happen and (2) sooner or later there will be one that is larger than the largest we have seen so far. Therefore, following the power-law logic, it will be no surprise if a large, established company fails in the coming years because of an out-of-control IT project.

Moreover, this research emphasizes the need to carefully assess the risk of new IT projects at the outset. One particular risk that managers must pay attention to involves technological components that are highly interdependent with other components. It may seem unlikely that one component can cause a substantial overrun. However, as demonstrated in our simulation, a single component can trigger cascading reactions in interconnected components, leading to substantial overruns. Thus, to minimize the risk of black swans [69], managers should exercise due diligence to identify riskier components (e.g., ones with high interdependencies) and provide additional resources to proactively manage them.

Finally, there exist several theory-based frameworks for evaluating the risk level of new IT projects, including options-based risk management [9, 11, 44] and an intelligent fuzzy approach [62]. Managers may benefit from using these frameworks to identify high-risk projects and prevent such projects from causing large cost overruns. Further, given the global nature of IT projects, managers may be able to mitigate technical risk by considering IT project teams' cultural composition [52].

## Conclusion

The extent to which an IT project is completed within budget is a key dimension of its performance and ROI. This research empirically examines and attempts to theoretically explain the probability distribution of IT project cost overruns. Using the largest dataset of its kind, our research shows that IT project cost overruns follow a power-law distribution, with a large number of small overruns and a smaller, but significant, number of very large overruns in a fat upper tail. Our research shows that IT projects have far greater cost risk than is commonly assumed. Incorrectly assuming a normal (Gaussian) or near-normal (e.g., log normal) distribution for cost risk, as is common, increases organizations' exposure to such risk by severely underestimating the probability of large cost overruns. Moreover, decision-making theory suggests this is important because ignoring fat-tailed, high-impact risk is associated with riskier, suboptimal decision making, which may lead to undertaking highly risky IT projects– projects that would not have been started had the true risk been known.

Our research contributes to IS research on IT project performance by highlighting both the need to examine project overrun data in greater depth (beyond mean and SD) and the need to advance our understanding of extreme project cost overruns that are often observed in the real world. In addition, we build on Simon's [65] conceptualization of sub-systems and interactions between sub-systems of large systems by synthesizing self-organized criticality and interdependency within the context of IT systems. Moreover, the power law distribution discovered in this research suggests that studying overruns requires large sample sizes, robust analytical methods, and caution when generalizing from small sample studies. Key practical implications follow from the predictive property of power laws: extreme overruns will occur that exceed previous observations, it is only a matter of time. This underscores the importance of high-quality planning and risk management when embarking on IT projects.

## Acknowledgment

The authors wish to thank the following for their highly useful comments on the earlier versions of the manuscript and/or on the ideas central to the paper: *JMIS* Editor-in-Chief Vladimir Zwass, three anonymous reviewers, Aaron Clauset, Henrik Flyvbjerg, Therese Graversen, Jens Schmidt, and Peter Sestoft.

## Disclosure Statement

No potential conflict of interest was reported by the authors.

## Notes on contributors

*Bent Flyvbjerg* is the first BT Professor and inaugural Chair of Major Programme Management at the University of Oxford's Saïd Business School, the Villum Kann Rasmussen Professor and Chair of Major Program Management at the IT University of Copenhagen's Computer Science Department, and Senior Research Fellow in Management at Oxford University's St. Anne's College. He holds a Ph.D. in Economic Geography from Aarhus University, Denmark, and two higher doctorates – Dr. Techn. and Dr.Scient. – in engineering and science, respectively, from Aalborg University, Denmark. Dr. Flyvbjerg's research focuses on decision making, project and program management, and risk, especially of big capital investments. He is the author or editor of 10 books and 200+ papers, translated into 20 languages. He has published in *Nature, Harvard Business Review, International Journal of Project Management, Energy Policy, World Development, Environmental Science and Policy, Oxford Review of Economic Policy*, and many others. He holds a knighthood and was awarded two Fulbright Scholarships, among other honours.

*Alexander Budzier* is Fellow in Management Practice in the Field of Information Systems at University of Oxford's Saïd Business school. He holds a DPhil in Management Studies from the University of Oxford and a Dipl. Wirtschaftsinformatik from TU Dresden. Dr. Budzier's research focuses on IT project management, general project management, and project risk. His work has been published in several scholarly journals, including *Journal of Information Technology, Energy Policy*, and *Environment and Planning A*.

*Jong Seok Lee* is an Assistant Professor in the Department of Accounting and Information Management in the Haslam College of Business and Economics at the University of Tennessee, Knoxville. He holds a Ph.D. in Computer Information Systems from Georgia State University. Dr. Lee's research focuses on behavioral and managerial issues in IT project management and software development. His work has been published in several scholarly journals, including *Information Systems Research, Journal of Management Information Systems, Journal of Association for Information Systems, European Journal of Information Systems*, and others.

*Mark Keil* is a Regents' Professor of the University System of Georgia and the John B. Zellars Professor of Computer Information Systems in the J. Mack Robinson College of Business at Georgia State University. He holds B.S.E., S.M., and D.B.A. degrees from Princeton University, M. I.T. Sloan School, and Harvard Business School, respectively. Dr. Keil's research focuses on IT project management and decision making and includes work on preventing IT project escalation, improving IT project status reporting, and identifying and managing IT project risks. He has published over 120 peer-reviewed papers. He serves on the editorial board of the *Journal of Management Information Systems* and has previously served in senior editorial roles for other journals including *MIS Quarterly, Information Systems Research*, and others. He is a past Division Chair for the Academy of Management.

*Daniel Lunn* was a Fellow at Worcester College and Professor of Statistics Emeritus at the Department of Statistics, University of Oxford. His research interests included reliability, survival analysis, multivariate analysis of high-dimensional data, directional data, and the analysis of clinical trials. He was particularly interested in the application of Bayesian methods. Dr. Lunn passed away before the completion of the paper's review process.

*Dirk W. Bester* studied actuarial science at the University of the Free State in Bloemfontein, South Africa, after which he received a Rhodes Scholarship to attend Oxford University. He completed a D. Phil. in Statistics under the supervision of David Steinsaltz, following which he worked as Chief Science Officer of Sciemus Ltd., calculating insurance risk of rockets, satellites, and power stations. He is currently working in the banking sector.